\documentstyle[12pt,aaspp4]{article}
\begin{document}

\def\plottwox#1#2{\centering \leavevmode
\epsfxsize=.5\columnwidth \epsfbox{#1}
\epsfxsize=.5\columnwidth \epsfbox{#2}}

\def\cm{{\,\rm cm}}
\def\kpc{{\,\rm kpc}}
\def\kms{{\rm km\,s^{-1}}}
\def\erg{{\rm erg}}
\def\ergs{\rm erg\,s^{-1}}
\def\phs{{\rm ph\,cm^{-2}s^{-1}}}
\def\Lx{L_{\rm X}}
\def\Lb{L_{\rm B}}
\def\fb{f_{\rm B}}
\def\fx{f_{\rm X}}
\def\fhx{f_{\rm HX}}
\def\foiii{f_{\rm OIII}} 
\def\NH{N_{\rm H}}
\def\arcm{$\arcmin$}
\def\arcs{\arcsec}
\def\keV{\,{\rm keV}}
\def\ergsec{\,{\rm erg\,s^{-1}}}
\def \eg           {{e.g.}}
\def \date         {\ifcase\month \message{zero} \or
                    January \or February \or March \or April \or May \or June 
                    \or July \or 
                    August \or September \or October \or November \or 
                    December \fi
                    \space\number\day, \number\year}

\def\Halpha{{\rm H}\alpha}
\def\Feka{{\rm Fe\, K}\alpha}
\def\Mgka{{\rm Mg\, K}\alpha}
\def\Sika{{\rm Si\, K}\alpha}
\def\Hbeta{{\rm H}\beta}
\def\Hgamma{{\rm H}\gamma}
\def\OI{[O\thinspace{\sc i}]\thinspace$\lambda$6300}
\def\OII{[O\thinspace{\sc ii}]\thinspace$\lambda$3727}
\def\OIII{[{\rm O}\,{\sc iii}]\thinspace\lambda{5007}}
\def\NeIII{[Ne\thinspace{\sc iii}]\thinspace$\lambda$3869}
\def\NII{[N\thinspace{\sc ii}]\thinspace$\lambda$6584}
\def\SIIone{[S\thinspace{\sc ii}]\thinspace$\lambda$6716}
\def\SIItwo{[S\thinspace{\sc ii}]\thinspace$\lambda$6731}
\def\SII{[S\thinspace{\sc ii}]\thinspace$\lambda$6716,6731}
\
\def\remark#1{{\bf (smao: #1)}}
\def\remarkxia#1{{(\bf #1 -- xia)}}

\def\pcmcu{\hbox{$\cm^{-3}$}}
\def\pcmsq{\hbox{$\cm^{-2}$}}

\title{CHANDRA OBSERVATIONS OF MRK 273: UNVEILING THE CENTRAL AGN 
AND THE EXTENDED HOT GAS HALO}

\author{X.Y. Xia$^{1,2}$, S.J. Xue$^{2}$, S. Mao$^{3}$, Th. Boller$^{4}$,
Z.G. Deng$^{5,2}$, and H. Wu$^{2}$}

\altaffiltext{1}{Dept. of Physics, Tianjin Normal University,
        300074 Tianjin, China}
\altaffiltext{2}{National Astronomical Observatories, 
                Chinese Academy of Sciences, A20 Datun Road, 100012 Beijing,
                China}  
\altaffiltext{3}{Univ. of Manchester, Jodrell Bank Observatory,
          Macclesfield, Cheshire SK11 9DL, UK}  
\altaffiltext{4}{Max-Planck-Institute f\"ur Extraterrestrische Physik,  
        Postfach 1312 D-85741 Garching, Germany}  
\altaffiltext{5}{Dept. of Physics, Graduate School,
        Chinese Academy of Sciences, 100039 Beijing, China}

\received{\date}
\accepted{ }

\begin{abstract}

We report X-ray observations of the field containing the ultraluminous 
IRAS galaxy Mrk~273. The data were
obtained using the ACIS-S3 instrument on board Chandra.
The high resolution X-ray image, for the first time, reveals
a compact hard X-ray nucleus in Mrk~273. Its position is coincident
with the northern nucleus identified in the optical, 
infrared, radio and in molecular CO maps.
Its X-ray energy distribution is well described by a 
heavily obscured AGN spectrum
with an absorbed power-law plus a narrow $\Feka$ emission line at 6.4\,keV.
The neutral hydrogen column density is about $4\times10^{23}\cm^{-2}$,
implying an absorption-corrected X-ray luminosity (0.1--10 keV) for the nucleus
of $\Lx\approx 6.5\times 10^{43}\,\ergs$ for
$H_0=50\,\kms {\rm Mpc}^{-1}$. The X-ray properties therefore firmly
establishes the northern
nucleus of Mrk~273 as a Seyfert 2 active galactic nucleus.
 
There are also bright soft X-ray clumps and diffuse soft X-ray emissions
surrounding the central hard X-ray nucleus within  the
$10\arcsec$ of the nuclear region. Its spectrum can be
fitted by a MEKAL thermal model with
temperature of about 0.8\,keV and high metallicity ($Z\sim 1.5Z_\odot$)
plus emission lines from $\alpha$ elements and ions. We find 
that a soft X-ray clump, about $4\arcsec$ (projected separation of about 4 kpc)
southwest of the northern hard X-ray
nucleus, is coincident with a nebula with strong $\OIII$ emissions. 
Further outside the central region, the Chandra observations reveal
a very extended hot halo in Mrk~273. The X-ray halo 
encompasses the entire optical tidal tail and plume, with a
projected diameter of about $108 \kpc\times 68 \kpc$.
The total soft X-ray luminosity (0.1--2.4\,keV) of the hot halo is 
$\Lx\approx 1.9\times 10^{41}\,\ergs$,
in the range of the soft X-ray luminosity of bright elliptical galaxies.
The temperature of the hot gas is about 0.62\,keV with 
a low metallicity ($Z \sim 0.1 Z_\odot$).
We discuss the nature of the AGN  in Mrk~273 and the
implications of our results on the origin of X-ray halos
in elliptical galaxies.

We also discuss the properties of Mrk~273x, a background AGN at
redshift $z=0.46$ in the Mrk~273 field. The AGN has an X-ray luminosity
of $\Lx \approx 2.43\times 10^{44}\ergs$ in the 0.5-10\,keV band.
Its X-ray properties resemble 
those of Seyfert 1 galaxies while its optical properties are similar to
Seyfert 2 galaxies. Such mixed classifications in the optical and X-ray 
may be a challenge for the unification scheme of AGNs.
\end{abstract} 


\section{INTRODUCTION}

Mrk~273 is an ultraluminous IRAS galaxy  (ULIRG) at redshift
$z = 0.0378$. It is a merging
galaxy that shows a striking long tidal tail ($\sim 1\arcmin$)
to the south, and a large tidal plume to the northeast 
(see Fig. \ref{fig:halo}). B-band images uncover
diffuse clumps in the northeast plume. These plumes
exhibit unusual optical spectra that are consistent with 
excitations by the shock plus precursor mechanism (Dopita \& Sutherland
1995; Xia et al. 1999).
Deep narrow-band $\Halpha$ and $\OIII$ images (Armus et al. 1990) reveal
filaments and arc structures
in the northeast plume and patchy areas of ionized gas (extending
tens of kpc) along the southern tidal tail. These
observations of the extended structures provide
evidence for shock excitations produced during merging. 

The nuclear region of Mrk~273 is extremely complex. It shows double
nuclei from both ground based images and HST optical (WFPC2),
K-band (NICMOS) images (Knapen et al.  1998;
Soifer et al. 2000; Carilli \& Taylor 2000). 
The projected separation between the two nuclei 
in the K-band is $1.1\arcsec$. The $\Hbeta$ and $\OIII$ maps obtained using 
integral field spectroscopy of Mrk~273 show that 
there are two distinct emission regions separated by $4\arcsec$
(Colina et al. 1999). The northeastern
region coincides with the optical and K-band nuclei; 
it is strong in the $\Hbeta$ emission but 
weak in the $\OIII$ emission,
i.e., it exhibits LINER spectral characteristics.
On the other hand, the southwestern region is dominated by diffuse
$\OIII$ emission and Colina et al. (1999) identified this region as a
Seyfert 2 nebula. Radio observations (Condon et al. 1991; Knapen et al. 1997) show that
there are three components: the northern and southwestern components
are coincident with the two K-band nuclei respectively 
while the southeastern component has a faint blue optical counterpart 
which was identified as a star-cluster by Scoville et al. (2000).
High resolution radio continuum and neutral hydrogen 21cm 
absorption observations by MERLIN, Very Large Baseline Array,
and Very Large Array 
(Cole et al. 1999; Carilli \& Taylor 2000) reveal a gas disk associated with
the northern nucleus. The disk has a diameter of $0.5\arcs$
with an inclination angle of $53^\circ$ and an average neutral
hydrogen column density $\sim 1.7\times10^{22}\,{\rm cm}^{-2}$. 
Furthermore, the high resolution CO(2-1) map by Downes \& Solomon (1998) 
uncovers a bright $0.35\arcsec \times 0.2\arcsec$ 
CO core in the nuclear molecular disk of the northern nucleus of Mrk~273. 
This is the most luminous extreme compact starburst region for local
ULIRGs with an infrared luminosity of $L_{\rm ir}\approx 6\times 10^{11}\,\L_{\odot}$
and a molecular mass of $1\times 10^{9} M_{\odot}$. 
In short, the Mrk~273 nuclear region is very complex and
it is not even clear how many different components there are,
let alone whether various components are powered by AGN and/or starbursts.

X-ray observations provide an independent probe of the physical processes
in merging galaxies such as Mrk~273. 
ASCA observation shows that there is
a heavily obscured active nucleus in Mrk~273 (Turner et al. 1997, 1998;
Iwasawa 1999). Hard X-ray emissions above 3\,keV and a narrow 6.4\,keV
$\Feka$ line were detected. An absorbed power-law model provides
an acceptable fit to the spectrum although
a reflection-dominated model can not be rejected either, based on the
ASCA data (Iwasawa 1999). 
However, given that ASCA's spatial resolution is $2\arcmin$, 
the position of the AGN cannot be pinpointed; this is particularly serious
because in the field of Mrk~273, ROSAT HRI images show that there is a
background X-ray source (Mrk~273x at redshift $z=0.46$) about $1.3\arcmin$ away (Xia et al. 1998).
So the ASCA data include contributions from both Mrk~273
and Mrk~273x. The sub-arcsecond resolution and 
high-energy sensitivity of Chandra are therefore 
essential for understanding the nature of X-ray emissions in
the Mrk~273 field.

Chandra observations are interesting for yet another important
reason. Elliptical galaxies are known to have hot gas halos; however,
the origin of the hot gas is still under 
debate (O'Sullivan et al. 2001; Sansom, Hibbard \& Schweizer 2000).
There is now firm evidence supporting the merger
scenario for the formation of elliptical galaxies, through 
major and/or multiple merger (e.g., Hernquist et al. 1996;
Borne 2000; Bekki 2001; Cui et al. 2001 and references therein).
Based on the ASCA data for 4 ULIRGs 
(Mrk\,231, Mrk\,273, Arp\,220 and NGC\,6240),
Iwasawa (1999) pointed out that the soft
X-ray emissions of these ULIRGs are thermal with a temperature 
of $(0.5-1)\times 10 ^{7}$ K. Incorporating
the extended morphology of soft X-ray emissions from
NGC 6240, Arp 220 and NGC 3690 observed by ROSAT HRI,
a thermal origin 
of the soft X-ray emission is acceptable. Furthermore, 
two-component fits to the soft X-ray data are better than 
one-component fits for these four ULIRGs and the low 
temperature component  is more extended
than the high temperature component (Iwasawa 1999).
A plausible explanation is that the high temperature component 
is from central starbursts and/or AGNs while the more extended low-temperature
component is provided by the shock-heated gas produced in mergers.
A high-resolution soft X-ray image is therefore necessary to
disentangle the X-ray emission from different parts of merging galaxies.

For the reasons described above, we have observed the
Mrk~273 field using the ACIS-S3 instrument on Chandra; in this paper we
report the results of these observations. 
Our focus is on Mrk~273 due to its implications for galaxy merging
and formation, although we do briefly discuss
the nature of the background source Mrk~273x (see \S\,5).
The structure of this paper is as follows. In \S\,2, 
we describe the observations  and data reduction.
In \S\,3, we present the images for Mrk~273 in both the soft
and hard X-ray band. These images reveal a compact hard X-ray nucleus and
several soft X-ray clumps accompanied by a very extended  hot halo.
In \S\,4, we study the spatially-resolved
spectral behaviors for the nuclear and extended X-ray emissions
in Mrk~273. 
In \S\,5, we discuss the X-ray results for Mrk~273x from the Chandra
observations.
And finally in \S\,6, we summarize and discuss our results,
particularly concerning the nature of the AGN in Mrk~273 and the connection 
to the origin of hot gas in elliptical galaxies.
Throughout this paper, we use a Hubble constant of
$H_0=50\,\kms {\rm Mpc}^{-1}$ and 
an Einstein-de Sitter ($\Omega_0=1$) cosmology although the latter has
little influence on our results due to the low-redshift ($z \sim
0.0378$) of Mrk~273. At the redshift of Mrk~273, $1\arcsec$
corresponds to 1.03\,kpc for the adopted cosmology.

\section{OBSERVATIONS AND DATA REDUCTION}

Mrk~273 and Mrk~273x were observed using the Chandra ACIS-S3 CCD chip
on 19 April 2000, with a total exposure time of $\sim$ 47ks. The observation 
was performed in the very faint (VFAINT) mode with ACIS working at 
the focal temperature of $-120^\circ$\,C. These two objects separated by
$\sim$ $1.3\arcmin$ were positioned on the same chip with offsets of $65.5\arcsec$
and $37.6\arcsec$ from the ACIS-S aim-point, respectively.

The data were initially processed with the
pipeline SDP version R4CU5UPD2,
available when the data were first available to us;
this pipeline has an astrometric error of about $3\arcs$
\footnote{http://asc.harvard.edu/mta/ASPECT/aspect\_caveats\_prior.html}.
The data used in this paper were reprocessed 
by the Chandra team using the new version of the software
(SDP R4CU5UPD13.2) in January 2001. The
most updated calibration files (such as ACIS QEU and gain 
map as well as observation-specific bad pixel lists)
were used in the new data products. The data were also filtered to
include only standard event grades 2, 3, 4 and 6 (see Chandra
Proposer's Observatory Guide, 2000).

The improvements in the processing software and calibrations lead to an
astrometric error of about $1.5\arcs$. However, this position
accuracy is still insufficient to align various nuclear components 
discovered in the optical, infrared and X-ray (see \S\,1).
By carefully examining both the Chandra image and
a WFPC2 HST image of Mrk~273, we found three 
point sources (shown in Fig. \ref{fig:smoothed} as A [i.e., Mrk~273x], B, C) 
that are present in both sets of images. These point 
sources allow us to determine the astrometry in the Mrk~273 field
to a much higher accuracy, $\sim 0.3\arcsec$;
this accuracy is crucial to register the X-ray nuclear components with
those in other wavebands.

The pre-processed and astrometrically-aligned
data products were then analyzed with the standard
CXC CIAO package (version 2.0). It begins with an inspection of 
bad aspect times and high background times. A background
flare event was detected at the beginning of the observation; the 
corresponding time intervals with large background rate (i.e. 20\% over 
the quiescent value) were thus removed, yielding an effective exposure 
time of $\sim41$ ks in the energy range 0.3--10\,keV. 
All the followup spatial and spectral analysis will be restricted to
this energy band; the April 2000 release
of the ACIS CCD calibration files (FEF) was used. 
As the standard CIAO tasks do not produce ARF and RMF for an
extended source, we also made use of the contributed software,
CALCRMF\footnote{http://hea-www.harvard.edu/\~\,jcm/asc/dist/av/av104.tar},
for this purpose.

\section{THE X-RAY IMAGE OF MRK~273}

Fig. \ref{fig:raw} shows the broad band (0.3--10keV) raw image of 
Mrk~273 and Mrk~273x at the full resolution (0.49$\arcs$/pixel) of Chandra ACIS-S3. 
It is clear from Fig. \ref{fig:raw} that the X-ray structure of Mrk~273 is
complex with a bright nuclear region and very extended diffuse emission.
Fig. \ref{fig:smoothed}
displays a true color image of Mrk~273, produced by
a mosaic of three X-ray bands, namely,
red: 0.3--1\,keV, green: 1--3\,keV and blue: 3--10\,\,keV.
Images in each of the three bands were adaptively smoothed
using the CIAO task {\it csmooth}, at 3$\sigma$ significant levels.
The smoothing function
we used is a scale-variable two-dimensional Gaussian with
a maximum scale of 10 pixels ($\approx 5\arcs$) to avoid inducing any
over-smoothing effects. 
The inset at the bottom left
corner is for the central $20\arcs \times 20\arcs$ nuclear region. 
The complex X-ray emission structures 
are clearer in this picture. It is obvious from Fig. 
\ref{fig:smoothed} that the hard X-ray (3-10\,keV) emission for
Mrk~273 is compact (the white central region)
while the soft X-ray emission (0.3--1\,keV) consists of diffuse 
halos and bright complex structures. Furthermore,
there are obvious intermediate X-ray band emissions surrounding
the hard compact region shown as yellow
in Fig. \ref{fig:smoothed}. We will discuss the possible origin
for these emissions in \S\,6.

Fig.~\ref{fig:energyEvolution} shows more clearly the evolution of the 
nuclear region of Mrk~273 with the X-ray energy. As can be seen
from this figure, the hard X-ray ($>$ 3\,keV) emission originates
from a compact region while the soft X-ray emission is diffuse 
with some bright clumps. The dominant soft X-ray bright clumps
are in the northeast and southwest. As X-rays become harder,
the northeastern clumps become brighter while 
the southwestern clumps become fainter. 
This indicates that the northeastern clumps are harder than the
southwestern clumps. Note that there is also a soft X-ray faint region
in the northwest direction which can be clearly seen in the
0.6-0.8keV and 0.8-1.1keV panels.

\subsection{THE NUCLEAR POSITION}

Fig.~\ref{fig:nuclearRegion} is an adaptively smoothed 0.3--1\,keV soft X-ray image 
of the nuclear region. Four concentric circles indicate the 
$1\arcsec, 3\arcsec, 6\arcsec$ and $10\arcsec$ regions around
the central nucleus. The square at the center of circles indicate
the position of the compact hard X-ray source.
It is clear that there are 4 soft X-ray bright clumps ,labelled as N, SE, SW1 and SW2,  
surrounding the hard X-ray source. 
The three plus signs in Fig.~\ref{fig:nuclearRegion} indicate the three resolved radio
components as given by Knapen et al. (1997).
Table 1 lists the positions of the 
hard X-ray component and soft X-ray bright clumps
together with the positions of the three radio components.
The absence of soft X-ray emission in the hard X-ray compact region
indicates that most soft X-ray emissions in the nucleus of Mrk~273 are
absorbed.

It is obvious from Fig.~\ref{fig:nuclearRegion} and Table 1 that the
hard X-ray position is coincident with the northern
radio component within the Chandra astrometric uncertainty
($\sim 0.3 \arcs$). This  northern component also has counterparts in the
K-band, mid-infrared and $\Halpha$\ ($\Hbeta$) narrow band images.
The compact hard X-ray source is also likely to be
embedded in a very dense, bright core of molecular gas
(about $0.3\arcsec \times 0.2\arcsec$) at the center of this nucleus
(Downes \& Solomon 1998). There is little doubt that
the northern radio component of Mrk~273 hosts an AGN. 
The optical spectrum of this
nucleus is of LINER-type (Colina et al. 1999; for more see \S\,5).
On the other hand, the southwestern and southeastern radio components
do not have any hard X-ray counterparts. The southwestern radio component 
is coincident with the southwestern K-band nucleus, which has been
proposed to be the second nucleus of Mrk~273. However, the lack of hard X-ray
emissions makes the interpretation less clear.

The left and right panels of Fig. \ref{fig:OIIIHalpha}
shows the 0.3--1\,keV soft X-ray contours 
overlaying on $\OIII$ and H$\alpha$ narrow band image for nuclear region of
Mrk~273, respectively. The $\OIII$ and H$\alpha$ narrow band image
are obtained by the 2.16m telescope of Beijing Astronomical Observatory.
The circle in the left and right panel of Fig. \ref{fig:OIIIHalpha}
indicates the hard X-ray position. It is clear from 
this figure that the southwestern 
soft X-ray bright clumps (SW1 and SW2)
are coincident with $\OIII$ nebula and the hard X-ray source is within
the northeastern H$\alpha$ bright clump, which is identified as the northern nucleus
in multi-wavelengths (see \S\,1). In fact, from Table 1 the
angular distance between the hard X-ray position and the SW2 component
is about $4\arcsec$, the same as the angular distance between the northern nucleus
and southwestern `Seyfert 2' nebula in Colina et al. (1999), which is
dominated by the $\OIII$ line emission.

\subsection{THE EXTENDED HOT GAS HALO}

The adaptively smoothed 0.3--1\,keV soft X-ray image
of Mrk~273 is shown in the left panel of Fig.~\ref{fig:halo}.
The maximum smoothing scale is $5\arcsec$
(10 pixels) and the contours shown in the middle and right panels
(see below) are 3, 5, 10, 20, 30, 50, 100 and 300$\sigma$ significance levels respectively.
Analysis indicates
that most statistically significant photons are inside 
in the outer ellipse in the left panel of Fig.~\ref{fig:halo}
and its major and minor radius are $52\arcsec$ and $33\arcsec$, respectively.
These correspond to a physical diameter of $108\kpc\times 68\kpc$.  This is
the largest extended soft X-ray halo that has been identified 
in an ultraluminous IRAS galaxy. Furthermore, as we have shown,
the high-resolution observation of Chandra allows us to compare fine structures with 
optical images at comparable resolution.
The middle panel of Fig.~\ref{fig:halo} shows
a deep R-band image of Mrk~273 (Armus et al. 1990) overlaid with the
soft X-ray contours.
It is obvious that the soft X-ray halo encompasses the
whole Mrk~273 optical image, including the southern long tidal
tail and northeast plume. In the right panel of Fig.~\ref{fig:halo},
the soft X-ray contours are overlayed on an $\Halpha$
narrow band image of Mrk~273 
obtained using the 2.16m telescope of Beijing Astronomical Observatory. 
We can clearly see from this plot that the southern and northeastern
soft X-ray structure are almost the same as the $\Halpha$  structure. 
Curiously, while the X-ray extension seems agree well with the $\Halpha$
extension, it shows a slight offset to the west of the optical tidal tail. 
This may be a natural consequence of the different behavior
of collisionless stars and collisional gas (outflow)
during galaxy merging as seen in Arp 299 (Hibbard \& Yun 1999). 
We will return to the origin of the extended halo in \S\,6.2.

\section{SPECTRAL ANALYSIS}

Given the sub-arcsecond resolution of Chandra, 
we can perform not only global but also
spatially-resolved spectral analysis. The latter is particularly
important for understanding the relative contribution of the
AGN/starburst contributions to the X-ray emission,
and by implication the origin of hot gas in the
descendents of mergers -- elliptical galaxies. 
High spatial and spectral resolution of Chandra also allow one to 
separate, for the first time, the contributions
to the hard X-ray emissions from Mrk~273 and Mrk~273x (see Fig. \ref{fig:raw});
this was impossible for ASCA due to its limited 
spatial resolution. 

\subsection{THE SPECTRUM OF THE CENTRAL $10\arcs$ REGION}

   From Figs. \ref{fig:energyEvolution} and \ref{fig:nuclearRegion} it is obvious  
that the 3-10\,keV hard X-ray emission of Mrk~273 is mainly concentrated
in the central compact region which shows little
soft X-ray emission. Therefore, to study how the X-ray emissions evolve
as one moves away from the central nucleus, we
extract spectra for the four annulus within $10\arcs$ 
region, as shown in Fig.~\ref{fig:nuclearRegion}, respectively.
The spectra were binned so that each bin contains at least 15
counts. The results are shown
in Fig. \ref{fig:ringSpectra}. It is clear from this figure
 that within the inner-most $1\arcs$
region, most photons are hard with energy higher than 3\,keV and there are
few photons with energy below 0.8\,keV.
As the radius increases,
the counts in soft X-ray band (0.3--2\,keV) increase significantly, but
the counts of hard X-ray photons have no substantial increase. More
specifically, more than 80\% hard X-ray photons are from the inner
1$\arcs$ region. and when the radius 
is increased to
$3\arcs$, the soft X-ray emission (0.3--2\,keV) already dominates
and there is no
big difference in counts within the $6\arcs$ and $10\arcs$ circles.
This result confirms the visual impression that the
hard X-ray emission is mainly from the central compact region while the
soft X-ray (below 2\,keV) contribution is from a more extended area.

Fig. \ref{fig:modelFit} is the spectrum for the whole 10$\arcs$ region of Mrk~273 
from 0.3 to 8\,keV, which is more reliably calibrated for the ACIS. 
The spectrum is clearly complex and consists of
at least three components. At energies above 3\,keV,
there is a heavily absorbed power-law plus
a narrow 6.4 keV $\Feka$ line. The other two components are
a less absorbed power-law and a MEKAL thermal plasma emission
with many evident line features, in particular Fe L  
and Ne line complex, and other $\alpha$ elements emission lines
(such as $\Mgka$, $\Sika$ lines, OVII and OVIII lines). 
The parameters of the
spectral fitting are given in Table 2 and the information for the emission 
lines is given in Table 3 and shown in Fig. \ref{fig:emissionLines}. These
components are discussed in more detail below.

It is clear from Fig. \ref{fig:modelFit} and Table 2 that the central hard X-ray
compact source has an absorbed power-law component and 
has a strong 6.4\,keV $\Feka$ emission line with an EW
of 213 eV. The neutral hydrogen column density 
is $N_{\rm H} \approx 4.1\times 10^{23}$\,cm$^{-2}$
and the photon index is $\Gamma\approx 2.1$. The absorption-corrected
X-ray luminosity is $\Lx\approx 6.5\times 10^{43}\,\ergs$,
which is two orders of magnitude below the far-infrared luminosity of Mrk~273.
It is worth comparing the value of neutral hydrogen column density determined by
ISO/SWS mid-IR observation, which is $N_{\rm H} \approx 5\times 10^{23}$\,cm$^{-2}$
(Genzel et al. 1998). The similarity of the $N_{\rm H}$ values obtained
using these independent methods gives
strong support to this spectral model.
These fitting results indicate that Mrk~273 contains an obscured moderate luminosity AGN. 
Incorporating the image information (\S3.1), the northern nucleus 
of Mrk~273 is a Seyfert 2 nucleus with a hard X-ray compact source 
embedded in a  bright compact molecular core.

On the other hand, the MEKAL thermal plasma model yields
a satisfactory fit to the energy distribution
of the hot gas surrounding the central nuclear region.
Its temperature is about 0.8\,keV, 
with $N_{\rm H} \approx 1.56\times 10^{21}$\,cm$^{-2}$ and
a metallicity of about $1.5Z_\odot$
(cf. Table 2). Also, there are remarkable Fe L, Ne line complexes, and
$\Sika$, $\Mgka$, OVII and OVIII emission lines that
are typical thermal excitation lines in the soft X-ray band. 
The absorption-corrected soft X-ray luminosity of the thermal
emission is $\Lx\approx 2.6\times 10^{41}\,\ergs$. The
properties of this soft X-ray thermal emission 
are very similar to the diffuse emission (excluding all the detected point 
sources) in the antenna galaxy NGC 4038/4039 (Fabbiano, Zezas \& Murray 
2001) and those in the prototypical starburst galaxy NGC 253 
(Strickland et al. 2000). Therefore, the thermal bright soft X-ray
emission within the central 10$\arcsec$ of Mrk~273 is very likely
due to massive starbursts.

There is also clearly a less absorbed second power-law component from the
spectral fitting as shown in Table 2 and in Fig. \ref{fig:modelFit}
(the dashed line). We first identified it as the
scattered light from the central nucleus since the fitting
power-law index $\Gamma$ ($\approx$1.9) is similar to that for the
heavily absorbed power-law component. However, most of 
the scattered light below 1\,keV from the AGN can not escape from the
very dense core, due to the high neutral hydrogen column density
in the nuclear region even outside the central torus 
($\NH \sim 1.7\times10^{22}\,{\rm cm}^{-2}$, Cole et al. 1999).
This suggests that the second power-law component
shown in Fig. \ref{fig:modelFit} may have other contributions
in addition to any directly scattered light from the central nucleus. 

This component may be partly contributed to by the 
X-ray binaries (XRBs) or supernova remnants. 
It has become increasingly clear that they
can make a significant contribution to the X-ray luminosity in
starburst galaxies. Examples include
NGC 253 (Strickland et al. 2000; Pietsch et
al. 2001), M82 (Kaaret et al. 2001), interacting galaxies
NGC 4038/4039 (Fibbiano et al. 2001), ULIRGs Arp 220
(Iwasawa et al. 2001) and perhaps even in
the X-ray faint elliptical and S0 galaxies (Blanton et al. 2001).
As Cappi et al. (1999) points out, 
the typical spectrum for the sum  of low-mass XRBs and high-mass XRBs
can be described by an exponentially cut-off power-law with index $0.5 <
\Gamma < 2.5$ which seems to describe the observed 
Chandra spectrum for NGC 4038/4039.
Therefore, it is plausible that the unresolved XRBs and supernova remnants
give contributions to the less absorbed second power-law component
in spectral fitting for the central 10$\arcs$ region of Mrk~273.

\subsection{THE SPECTRUM OF THE EXTENDED HOT GAS HALO}

The spectral fitting for the hot gas halo has been performed 
for the region between the outer ellipse and the inner circle shown in the left panel
of Fig. \ref{fig:halo} excluding all bright point sources (shown as small circles)
in this region. Fig. \ref{fig:haloSpectrum} gives the MEKAL thermal model fitting 
and the fitting parameters are listed in Table 4. A thermal model with
temperature of 0.62\,keV, $N_{\rm H} \approx 3.0\times
10^{20}$\,cm$^{-2}$ and metallicity of about $0.1Z_\odot$ yields an
acceptable fit. The model dependent soft X-ray luminosity is
$\Lx\approx 1.9\times 10^{41}\,\ergs$.
Comparing with the
central soft X-ray spectral properties, the temperature of the extended hot
gas is clearly lower than that of the central
starburst region. However, the difference is not large.
The most striking difference between the central soft X-ray and the
extended hot gas is the metallicity. 
The central starburst region has metallicity higher 
than the solar metallicity while the more extended hot gas outside
seems to have a very low metallicity. We caution that we have 
modelled the halo as a single MEKAL thermal model; in reality, the gas
may be clumpy and have complex temperature structures. A map with higher 
signal-to-noise ratio will be very useful to further strengthen
our conclusions on metallicities. We briefly return to this important question 
of chemical enrichment in merging galaxies in \S6.2.

\section{MRK~273X}

Based on its optical spectrum, Mrk~273x is classified as a Seyfert 2 galaxy. ROSAT
observations indicate that it has an X-ray luminosity,
$\Lx\approx 1.1\times 10^{44}\,\ergs$ in the 0.1-2.4\,keV band,
one of the highest in the soft X-ray band among Seyfert 2 galaxies (Xia et al. 1999).
Borne et al. (1999) used HST I-band images of the
Mrk~273 field to analyze the luminosity distribution of faint galaxies 
surrounding Mrk~273x. They suggest
that Mrk~273x is the brightest elliptical galaxy in a relatively poor cluster. 
Hard X-ray observations have been carried out by ASCA and
BeppoSax. However, due to
their limited spatial resolution they can not resolve
Mrk~273x from Mrk~273. Therefore
the spectrum shown by Iwasawa (1999) is a composite spectrum
of Mrk~273 and Mrk~273x. The high resolution and high-energy response of
Chandra allow us to investigate the
X-ray properties of Mrk~273 and Mrk~273x separately and also test 
the variability for both Mrk~273 and Mrk~273x.

\subsection{THE SPECTRUM OF MRK~273X}

As can be seen from Figs. 1 and 2,
Mrk~273x is a bright X-ray point source. Figure 11 shows the observed
Chandra spectrum of Mrk~273x from 0.3 to 8\,keV. 
A power-law model with photon index 
$\Gamma=1.66^{+0.15}_{-0.11}$ and $N_{\rm H}=1.41^{+0.55}_{-0.50}
\times 10^{21}$\,cm$^{-2}$ 
plus a thermal component with temperature of $0.56^{+0.24}_{-0.24}$
 keV and solar abundance
give the best fit to the spectrum of Mrk~273x (the $\chi^2$ per degree
of freedom is 41.8/56). The absorption corrected
X-ray luminosity in the 0.5-10 keV band is 
$\Lx\approx 2.43\times 10^{44}\,\ergs$. The luminosity of the 
thermal component is $\Lx\approx 6.0\times 10^{42}\,\ergs$, about
3\% of the total X-ray luminosity of Mrk~273x. Since
Mrk~273x is an elliptical galaxy from its
de Vaucouleurs surface brightness profile (Borne et al. 1999),
the thermal component of X-ray
emission may naturally arise from the Mrk~273x host 
galaxy. From the spectral fitting, there is one significant emission line at 2.66 keV
(S\,XVI line) with EW$\approx 89\,{\rm eV}$. In contrast,
the $\Feka$ emission line is not convincingly detected 
(the EW upper limit for a line at 6.4\,keV is 30\,eV at the 90\% confidence level).

\subsection{THE VARIABILITY OF MRK~273 AND MRK~273X}

We reported that there was no significant variability for either Mrk~273 
or Mrk~273x during ROSAT observations (Xia et al. 1998). 
The timing analysis for Mrk~273 and Mrk~273x based on the Chandra data alone
also does not reveal any significant variability. In order to test for 
possible long term variabilities, we compile in Table 5
the X-ray fluxes of Mrk~273 and Mrk~273x in various X-ray bands
from the literature.

    From Table 5, the flux differences in the soft X-ray band (0.1-2.4 keV)
are about 33\%, 34\% between ROSAT and Chandra observations over a span of
8 years for Mrk~273 and Mrk~273x, respectively. In the 0.5-10\,keV band,
the difference for the total flux of Mrk~273 and Mrk~273x between 
the Chandra observation in 2000 and the ASCA observation in 1996
is about 21\%.
In the 2-10\,keV band, there are three reported observations for the
total flux. While the  Chandra and ASCA fluxes are within
30\% of each other, the BeppoSax flux is lower by about a factor of 2
(Risaliti et al. 2000).
To summarize, all the data except that from BeppoSax are consistent
with no significant variability. However,
much denser time sampling is required to firmly
establish the variabilities of Mrk~273 and Mrk~273x.
 
\subsection{THE NATURE OF MRK~273X}

The X-ray properties of Mrk~273x can be compared with those of
QSOs, Seyfert 1 and Seyfert 2 galaxies 
by ASCA and BeppoSax (e.g., Reeves \& Turner, 2000; Pappa et al, 2001). 
Although Mrk~273x is classified optically as a Seyfert 2 galaxy, its
X-ray properties are typical for a Seyfert 1 galaxy.
Its high X-ray luminosity of $\Lx\approx 2.43\times 10^{44}\,\ergs$ falls
into the regime of Seyfert 1 galaxies. Furthermore,
similar to many Seyfert 1 galaxies,
it is dominated by a power-law spectrum with
$\Gamma\approx 1.68$, there is no cutoff at low energies
and the $\Feka$ line is not detected.

Objects with Seyfert 2's optical spectrum but with Seyfert 1's
properties in the X-ray have been reported 
(for example NGC3147, Ptak et al. 1996; NGC7590, Bassani et al. 1999;
NGC4698, Pappa et al, 2001; NGC7679, della Ceca et al, 2001).
Mrk~273x is another such example, although the X-ray luminosity
of Mrk~273x and the ratio of the hard X-ray luminosity (in the 2-10\,keV band)
to the OIII luminosity,
 $\fhx/\foiii$,  are one of the highest even among this subclass of AGNs
(the $\fhx/\foiii$ ratio for Mrk~273x is 86). Given that these
sources have weak or no $\Feka$ detection and high $\fhx/\foiii$ ratios,
they could not be Compton thick objects
(with $N_{\rm H} > 10^{24}$\,cm$^{-2}$) because for such objects,
the EW of $\Feka$
is well above 1\,keV and the $\fhx/\foiii$ ratio is less than 1 
(Gilli et al. 1999). 
Therefore, the power-law spectrum of this type of object is not from reflection, 
but could be due to the lack of intrinsic absorption. 
The most straightforward
explanation for this type of object is that it lacks broad line regions and the
Seyfert 2 optical spectrum is intrinsic, but not due to heavy absorption
(Pappa et al. 2001). The existence of such objects challenges
the standard AGN unification model. Moreover, Mrk273x has a very high
X-ray to optical B-band flux ratio ($\approx 7$, Xia et al. 1998), which is 
only achieved by BL Lacertae objects (Stocke et al. 1991). The
strong optical emission lines
and the lack of dramatic X-ray variability (see \S5.2), however, do not support
Mrk~273x being a BL Lacertae object. Therefore, Mrk~273x still remains
a mysterious object that warrant further studies.

\section{SUMMARY AND DISCUSSION}

Compared with earlier satellites, the Chandra ACIS-S3 observation
of the Mrk~273 field unveils unprecedented details of the X-ray emissions
in the  merging galaxy Mrk~273.
These observations raise a number of important questions, which we will
discuss below.

\subsection{THE NUCLEAR REGION OF MRK~273}

Multi-wavelength high resolution imagings for the
nuclear region of Mrk~273 reveal intricate and puzzling details of 
the heavily dust-obscured nuclear region of Mrk~273 (see \S1 and \S3.1).
Some components in one waveband
have no clear counterparts in other wavebands. 
The northern K band putative nucleus has radio, mid-infrared and
$\Halpha$ narrow band counterparts. Combined with the fact that
there is a bright molecular core at the center, this nucleus most likely 
hosts a central AGN. This is further supported by the Chandra X-ray
observations, where we have identified a hard X-ray source coincident
with this northern nucleus within the astrometric uncertainty. 
The X-ray spectrum of the hard X-ray source is typical for 
Seyfert 2 galaxies.

The `contradiction' is from the comparison with the 
integral field spectroscopy of Mrk~273 by Colina et al. (1999) from which
the northern nucleus of Mrk~273 has a LINER optical spectrum. However,
as suggested by Terashima et al. (2000), a LINER 1 spectrum can also be
produced by photoionizations due to hard photons from a low-luminosity
AGN. The northern nucleus of Mrk~273 may be similar, although in this case,
the low-luminosity is not intrinsic but due to the heavy
attenuation of gas. The high neutral hydrogen column-density 
($\NH \sim 4.1\times10^{23}\,{\rm cm}^{-2}$) implies that
the soft X-ray photons could not escape from the central region, and thus
the photoionizations can only be due to hard X-ray photons from the Mrk~273 
northern nucleus. Therefore, a LINER 1 spectrum can also be a result of
photoionizations by hard photons from a high-luminosity AGN that is
heavily absorbed by a dense neutral gas. This mechanism 
may be common for ULIRGs where both central AGNs and large amount of
neutral gas are present in these merging galaxies.

On the other hand, the southwestern soft X-ray clumps are coincident
with a bright $\OIII$ nebula with angular distance of 4$\arcsec$ 
(projected separation of about 4 kpc) from the
northern nucleus (see Fig. \ref{fig:OIIIHalpha}). This region has a
spectrum dominated by the $\OIII$ line emission and is classified as 
a Seyfert 2 nebula by Colina et al. (1999). The northern nucleus
has an associated extended gas disk with an inclination angle of $53^\circ$
(Carilli \& Taylor 2000). The CO molecular disk has a
similar inclination angle from the velocity map of Downes \& Solomon (1998).
This orientation, combined with the relative 
positions of the hard X-ray compact source, 
bright $\OIII$ nebula, southwestern radio emission (Cole et al. 1999) as
shown in Figs. 4 and 5 and the narrow profile of $\OIII$ line
(${\rm FWHM}<260\kms$) (Colina et al. 1999),
 suggest that the southwestern bright $\OIII$ nebula may be 
an extended narrow line region: the $\OIII$ emission region 
may reside in an ionization cone centered on the hard X-ray source.
This geometry and the strong correlation between the soft X-ray clumps and
the high excitation (e.g. $\OIII$) optical line emission is similar
to the case in the nearby Seyfert 2 galaxy NGC~1068
(Young et al. 2001). 
Therefore, the photoionization source for both the $\OIII$ and soft
X-ray may be the central AGN associated with the northern nucleus. It is,
however, still possible that the excitation can be produced by
shocks occurring in galaxy merging (Xia et al. 1999).

\subsection{THE ORIGIN OF HOT GAS HALO OF MERGING GALAXY}

Figs. \ref{fig:smoothed} and 
\ref{fig:nuclearRegion} clearly show that the soft X-ray (0.3--2\,keV) image
of Mrk~273 is complex with several bright clumps 
surrounding a hard X-ray compact nucleus, irregular structure
 along the direction of southern tidal tail and
faint structure along the northern plume.
It extends beyond the optical emission region
with a projected diameter
of about $108 \kpc\times 68\kpc$.
A MEKAL thermal model with temperature of 0.62\,keV and metallicity of 
about $0.1Z_\odot$ provides the best fit. The model-dependent soft
X-ray luminosity of this hot gas halo is 
$\Lx\approx 1.9\times 10^{41}\,\ergs$. From Tables 2 and 4, the hot gas
properties in the central starburst region and the faint
very extended halo are very different. The very low metallicity 
implies that the diffuse hot gas halo has not yet been
chemically contaminated by stellar processes such as superwinds 
which are expected to carry a substantial amount of heavy metals
(Heckman 2001).

The temperature and luminosity of this very extended hot gas
halo are in the range of bright elliptical galaxies
(Brighenti \& Mathews 1997) and a sample of merging galaxies studied by
Read \& Ponman (1998). 
The X-ray data therefore suggest that merging can produce
hot halos that resemble those seen in elliptical galaxies, which
adds another piece of evidence for elliptical galaxy formation via mergers.
The low-metallicity of the gas indicates that the cooling time of the
gas may have been under-estimated by a factor of a few 
(e.g., Fig. 9-9 in Binney \& Tremaine 1987) since many previous
studies assumed solar-metallicities (e.g., Sarazin 1990).
This hot gas may  remain X-ray luminous for more than 
$\sim$ 1 Gyr, long after the stellar components at the center have
relaxed dynamically. The low metallicity, although somewhat uncertain,
suggests that part of the X-ray emission in elliptical galaxies may 
be due to secondary infalls in the pre-merger (group) environment
(e.g., Brighenti \&  Mathews 1998, 1999), which have not yet been
contaminated by the central starbursts.
This is also consistent with the recent work by
Bekki (2001) who studied multiple galaxy mergers using numerical simulations.
He showed that metals produced and ejected
in central star formation regions are mostly
mixed with the local interstellar medium.
Consequently, the ISM of the outer part of the merger is
less metal-enriched. The radial gradient in metallicity seen in Mrk~273
is certainly consistent with the picture, where
the 10\arcs\, central region has a
metallicity ($Z \sim 1.5 Z_\odot$) while the very extended hot gas halo 
has a much lower metallicity ($Z \sim 0.1 Z_\odot$).
We caution, however, that the star formation treatment in numerical 
simulations is not realistic in the sense that the multi-phase medium nature of
the interstellar medium is not properly taken into account. As a
consequence of this inadequacy,
the region that are chemically enriched may be under-estimated.

To conclude, the high resolution observation of the Mrk~273 field by
Chandra yielded important results for both Mrk~273 and Mrk~273x. For
Mrk~273x,  Chandra observation confirm that its X-ray properties 
are similar to Seyfert 1 galaxies while it is classified as a Seyfert 2
galaxy optically.
For Mrk~273, where our primary interests lie, Chandra observation,
for the first time, reveals a compact hard X-ray nucleus inside
a much more extended halo. Although the nuclear region of Mrk~273 
is complex, it is a typical Seyfert 2 nucleus with extended
narrow line region. The high temperature and high-metallicity thermal emission
component from the spectral fitting are obviously from the starburst region, 
while the much more extended hot gas
halo has a much lower metallicity and may have a different origin. The data
provide further support for the evolutionary connection between galaxy merging
and elliptical galaxies. More high-resolution deep images of
merging galaxies at various stages will be very useful 
for providing further clues of X-ray halos in merging galaxies and
their descendents, elliptical galaxies.

\acknowledgments

We are grateful to the CXC team for assistance in observations and
data reduction. We also thank Drs. H.J. Mo, A. Pedlar, J. Skelton,
Y. Gao, M.S. Yun, X.P. Wu, D. Wang, T.G. Wang 
and X.W. Liu for advice and helpful discussions.
Thanks are also due to Dr. Z.J. Jiang for help in the data reduction. 
This project was supported by the NSF of China and NKBRSF G19990754.
SM gratefully acknowledges a travel grant awarded by the NSF of China.

\clearpage


\figcaption[fig1.ps]{\label{fig:raw}
Broad-band (0.3--10keV) raw image of Mrk~273 and Mrk~273x 
at full resolution (~0.49$\arcsec$/pixel) of Chandra ACIS-S3. 
North is up and east is to the left.
The northeast bright source is Mrk~273x. The field of view
is $3.7\arcmin\times 3.7\arcmin$.
}

\figcaption[fig2.ps]{\label{fig:smoothed}
Adaptively smoothed true color image of Mrk~273,
with a maximum smoothing scale of 10 pixels ($\approx 5\arcsec$).
The image is produced by a mosaic of three X-ray bands, namely
the 0.3--1\,keV (red), 1-3\,keV (green) and 3-10\,keV (blue) energy ranges.
The inset shows a magnified view of the central region. Three point
sources (labelled as A, B and C) are used to register the Chandra
astrometry (see \S2). Object A is Mrk~273x.
}

\figcaption[fig3.ps]{\label{fig:energyEvolution}
The central $24\arcsec\times 24\arcsec$ region of Mrk~273 in
six X-ray bands; the energy range is indicated at the bottom left
of each panel. Notice how different regions evolve as the energy band
changes. It is apparent that the hard X-ray component is relatively compact.
}

\figcaption[fig4.ps]{\label{fig:nuclearRegion}
The soft X-ray image (0.3--1\,keV) of the nuclear region 
of Mrk~273. The square is for the hard X-ray
position and three crosses are for the three resolved
radio emission components (Knapen et al. 1997). The coordinates
for all these components are shown at Table 1. The hard X-ray position
is coincident with the northern radio component within the uncertainty
of the Chandra astrometry; this component 
also has counterparts in the K-band, mid-infrared and molecular CO maps.
The spectra for the emission contained within the four circles are shown 
in Fig. \ref{fig:ringSpectra}.
\label{fig4}}

\figcaption[fig5.ps]{ \label{fig:OIIIHalpha}
The left panel shows the $\OIII$ image overlaid by soft X-ray (0.3--1 KeV) contours while the right panel shows
the $\Halpha$ image overlaid by soft X-ray (0.3--1 KeV) contours for nuclear region of Mrk~273. In both panels, 
the center of the circle indicates the hard X-ray position.
}

\figcaption[fig6.ps]{ \label{fig:halo}
Adaptively smoothed X-ray image in the 0.3--1\,keV energy range
of Mrk~273. The maximum smoothing scale is 5\arcs.
The left panel is a smoothed image.
Notice that most photons must be included in the
outer ellipse which has a size $52 \arcsec \times 33\arcsec$ 
and a position angle of $13^\circ$.
The spectrum for the extended hot gas halo is extracted 
for the region between the outer ellipse and the inner circle, and is
shown in Fig. \ref{fig:haloSpectrum}.
The middle and right panels
are the R band deep image and H$\alpha$ narrow band image of Mrk~273 
overlaid by the Chandra soft X-ray contours
at 3, 5, 10, 20, 30, 50,100 and 300$\sigma$, respectively.
It is clear from these plots that the hot gas halo of Mrk~273 encompasses both the long 
tidal tail to the south and the plume to the northeast. Notice that
 the soft X-ray has almost the same structure as H$\alpha$ in the southern and 
northeastern direction, but offset with southern optical tidal tail. 
}

\figcaption[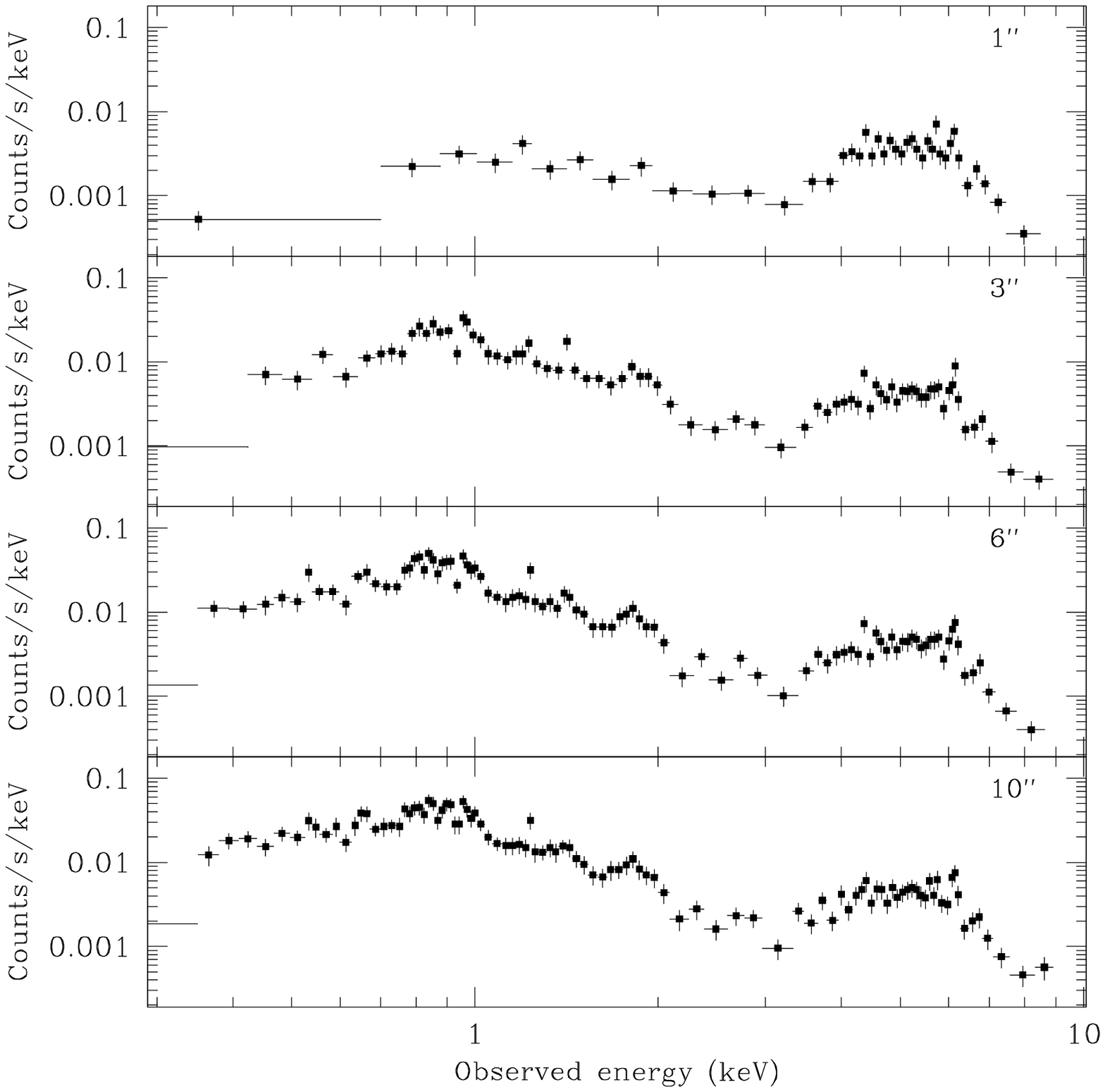]{\label{fig:ringSpectra}
The spectra for the central $1\arcs, 3\arcs, 6\arcs$ and $10\arcs$ 
nuclear regions.
} 

\figcaption[fig8.ps]{ \label{fig:modelFit}
The 0.3--8\,keV spectrum of the central 10\arcs\, region of Mrk~273. 
The spectrum is fitted well with the superposition
of a heavily absorbed power-law, a less absorbed
power-law, a thermal MEKAL component
plus 6.4\,keV $\Feka$, Fe L, Ne line complex, 
and other $\alpha$ elements emission lines. 
The fitting parameters are shown in Tables 2-3.
\label{fig6}
}

\figcaption[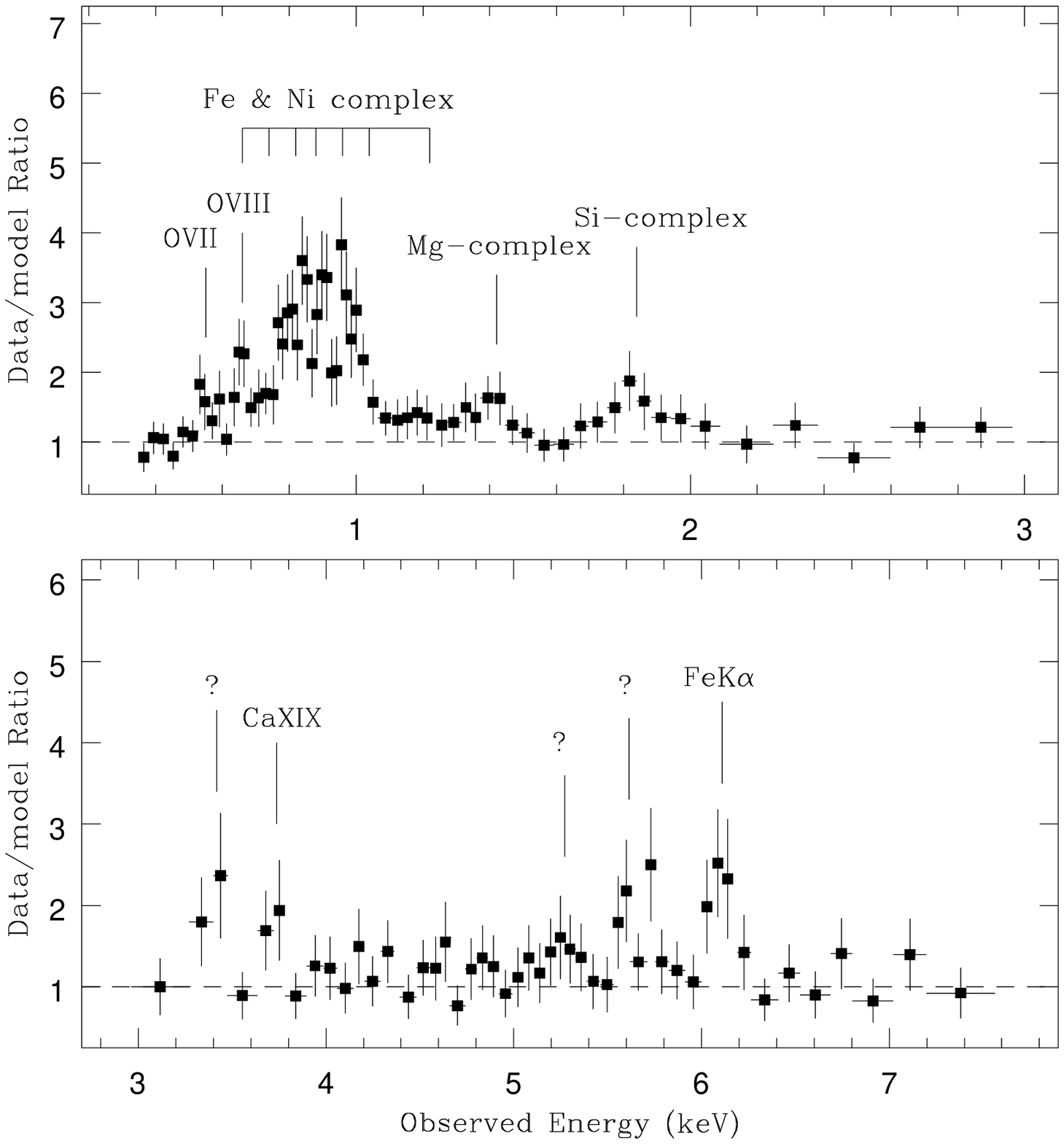]{ \label{fig:emissionLines}
The ratio of the data to the model continuum, highlighting
the features due to emission lines.
}

\figcaption[fig10.ps]{\label{fig:haloSpectrum}
The 0.3--2keV spectrum for the
extended hot gas halo for the region between the outer ellipse and the
inner circle in the left panel of Fig.~\ref{fig:nuclearRegion}.
A MEKAL model with temperature of 0.62\,keV and metallicity of
about $0.1Z_\odot$ provides a satisfactory
fit. The fitting parameters are given in Table 4.
}

\figcaption[fig11.ps]{\label{fig:mrk273x}
The observed 0.3--8\,keV spectrum for Mrk~273x. 
The spectrum is fitted well with the superposition
of an absorbed power-law, a thermal MEKAL component
and a line emission at 2.66keV (see \S 5.1 for the fitting parameters). 
}

\clearpage

\begin{deluxetable}{llll}
\tablecaption{Positions of the nuclear hard X-ray source, soft X-ray 
clumps and radio components}
\tablehead{
\colhead{Name} & \colhead{RA} & \colhead{DEC} & \colhead{Separation}
}
\startdata
Hard X-ray     &13:44:42.12 &55:53:12.9 &0.28\arcs   \nl  
Clump N        &13:44:42.11 &55:53:14.4 &1.22\arcs   \nl
Clump SE       &13:44:42.33 &55:53:11.5 &2.46\arcs   \nl
Clump SW1      &13:44:41.93 &55:53:11.0 &2.69\arcs   \nl
Clump SW2      &13:44:41.70 &55:53:11.5 &3.90\arcs   \nl
Radio N        &13:44:42.1171 &55:53:13.182 &0\arcs  \nl
Radio SE       &13:44:42.1677 &55:53:12.496 &0.81\arcs  \nl
Radio SW       &13:44:42.0372 &55:53:12.144 &1.24\arcs  \nl
\tablecomments{
Chandra corrected positions for various
nuclear components of Mrk~273 together with the 
the radio nuclear components from Knapen et al. (1997). 
The Chandra astrometry 
has been corrected using the three point sources shown in
Fig. \ref{fig:smoothed} and has an uncertainty of $\sim 0.3\arcsec$ (the
Chandra components are shown in Fig.~4).
The radio astrometry has an error of 2\,mas for the N and SE components and
27\,mas for the SW component (see Knapen et al. 1997, Table 2).
The K band northern nucleus is at the same
position as the radio northern (N) nucleus
 and the southwest K band nucleus is about $1\arcs$ to 
the northern nucleus, nearly coincident with the radio SW component. 
Column 1 is the name of the Chandra or radio components.
The second and third columns are the RA and DEC for the components
while column 4 is the separation between each component and the northern (N)
radio component.
}
\enddata
\end{deluxetable}

\begin{deluxetable}{ccccccccccccl}
\tiny
\tablewidth{0pt}
 
\tablecaption{Spectral fits to the central $10\arcsec$  region}
\tablehead{
\multicolumn{3}{c}{MEKAL\tablenotemark{a}} &&
\multicolumn{2}{c}{PL$_2$\tablenotemark{b}} &&
\multicolumn{5}{c}{PL$_1$\tablenotemark{c} + Line} & \nl
\cline{1-3} \cline{5-6} \cline{8-12} \nl
$\NH$ & kT & Z && $\NH$ & $\Gamma_1$ & & $\NH$ & $\Gamma_2$ &
E$_{\rm line}$ & $\sigma$ & EW & $\chi^2_\nu$ \nl
 [$10^{21}$\pcmsq]& [keV] & [$Z_\odot$] & & [$10^{21}$\pcmsq]& & &
 [$10^{23}$\pcmsq]& & [keV] &[keV]  & [eV] & \nl }
\startdata
$1.56^{+1.29}_{-1.56}$ & $0.77^{+0.09}_{-0.04}$ & $1.48^{+3.10}_{-0.25}$ &&    
$0.69^{+0.31}_{-0.22}$ & $1.94^{+0.31}_{-0.28}$ && $4.13^{+0.42}_{-0.58}$ &
$2.10^{+0.22}_{-0.25}$ &
$6.37^{+0.04}_{-0.05}$ & $2.9^{+5.1}_{-2.9}$ & $213^{+110}_{-94}$ &
{0.97} \nl
\enddata
\tablenotetext{a}{Emission measure as absorption corrected luminosity,
$L_{\rm MEKAL}(0.1-10\keV) = 2.6^{+0.32}_{-0.37}\times 10^{41}\ergsec$.}
\tablenotetext{b}{Emission measure as absorption corrected luminosity,
$L_{\rm PL_2}(0.1-10\keV)=1.5^{+0.40}_{-0.62}\times 10^{42}\ergsec$.}
\tablenotetext{c}{Emission measure as absorption corrected luminosity,
$L_{\rm PL_1}(0.1-10\keV)=6.5^{+1.8}_{-2.8}\times 10^{43}\ergsec$.}
\tablecomments{
Model: $\rm Absorption\times MEKAL(kT,Z)+Absorption\times PL(\Gamma_2)+
Absotption\times[PL(\Gamma_1)+Line]$, where 
MEKAL is for the Mewe, Kaastra \& Liedahl thermal plasma model, 
PL$_1$ \& PL$_2$ are for the two power-law models, and the emission
lines have Gaussian profiles.
}

\end{deluxetable}

\begin{deluxetable}{cccc}
\tablecaption{Line detections from the spectrum of  the
central $10\arcs$ region}
\tablehead{
\colhead{Rest energy} & \colhead{Line Flux} & \colhead{EW} & \colhead{Line} \\
\colhead{[keV]} & \colhead{[$10^{-6}\ \rm phs\ cm^{-2}\ s^{-1}$]} & \colhead{[eV]} & }
\startdata
6.37$^{+0.04}_{-0.05}$ & 6.2$^{+3.2}_{-3.0}$ &    213 & $\Feka$ \  \nl
5.81$^{+0.09}_{-0.05}$ & 2.7$^{+2.3}_{-2.3}$ &    126 & ?          \nl
5.47$^{+0.07}_{-0.16}$ & 0.4$^{+1.9}_{-1.9}$ &    100 & ?          \nl
3.88$^{+0.05}_{-0.06}$ & 0.5$^{+0.6}_{-0.5}$ &    110 & Ca XIX 	   \nl
3.55$^{+0.05}_{-0.06}$ & 0.6$^{+0.5}_{-0.6}$ &    195 & ?          \nl
1.91$^{+0.05}_{-0.04}$ & 1.0$^{+0.4}_{-0.6}$ &    110 & $\Sika$    \nl
1.49$^{+0.05}_{-0.04}$ & 1.0$^{+0.4}_{-0.7}$ &    75  & $\Mgka$    \nl
0.76-1.26\tablenotemark{a} &   -       &   -    & Fe \& Ni complex \nl 
0.69$^{+0.04}_{-0.04}$ & 2.5$^{+1.3}_{-1.0}$ &    65  & O VIII     \nl 
0.57$^{+0.17}_{-0.19}$ & 1.7$^{+2.0}_{-1.2}$ &    34  & O VII      \nl
\tablenotetext{a} {Thermal emission bump with many unresolved lines.}
\enddata
\end{deluxetable}

\begin{deluxetable}{ccccccc}
\tablecaption{Spectral parameters for the extended hot halo}
\tablehead{
\colhead{Model} & \colhead{$\NH$} & \colhead{kT} & \colhead{Z} & 
\colhead{$\chi^{2}$/d.o.f} & \colhead{$A_{\rm MEKAL}$\tablenotemark{a}} 
& \colhead{$\Lx$\tablenotemark{b}} \\
   & \colhead{[$10^{21}\ {\rm cm}^{-2}]$} & \colhead{[keV]} 
   & \colhead{[$Z_\odot$]}    & & \colhead{[$10^{-5}$]} 
   & \colhead{[$10^{41}\ergs$]} }  
\startdata
MEKAL & $0.30^{+0.71}_{-0.30}$ & $0.62^{+0.07}_{-0.13}$ 
      & $0.11^{+0.10}_{-0.06}$ & 20/27 & $3.8^{+1.4}_{-1.9}$ 
      & $1.9^{+0.5}_{-0.6}$ \nl   
\tablenotetext{a} {The model normalization, 
  $A_{\rm MEKAL}$, is defined as $10^{-14} \int n_{\rm e} n_{\rm H} dV / (4 \pi
(1+z)^2D_A^2)$, where
  $n_{\rm e}$ is the electron number density (\pcmcu), $n_{\rm H}$ is the hydrogen
number density 
  (\pcmcu) and $D_A$ is the angular diameter distance to the source in
	units of cm.}
\tablenotetext{b} {The luminosity in the 0.1-2.4\,keV band, corrected
for both the Galactic  and intrinsic absorptions.}
\enddata
\end{deluxetable}

\begin{deluxetable}{llllll}
\tablecaption{X-ray fluxes in units of $10^{-13}{\rm erg\,s^{-1}\,{cm}^2}$
for Mrk~273 and Mrk~273x\tablenotemark{a}}
\tablehead{
\colhead{Source} & \colhead{f(0.1-2.4keV)} & \colhead{f(2-10 keV)} & \colhead{f(0.5-10 keV)}& \colhead{Instrument} & \colhead{time}
}
\startdata
Mrk~273          &1.25 &     &     & ROSAT   &1992   \nl
Mrk~273          &0.94 &7.93 &8.72 & Chandra &2000   \nl
Mrk~273x         &1.10 &     &     & ROSAT   &1992   \nl
Mrk~273x         &0.82 &1.19 &1.85 & Chandra &2000   \nl
Mrk~273+Mrk~273x &2.35  &     &     & ROSAT    &1992  \nl 
Mrk~273+Mrk~273x &1.7\tablenotemark{b}  &7.0  &8.7  & ASCA    &1996  \nl 
Mrk~273+Mrk~273x &     &3.50 &     & BeppoSax&1998  \nl
Mrk~273+Mrk~273x &1.76 &9.12 &10.57 & Chandra &2000  \nl
\tablenotetext{a}{The flux by ROSAT, ASCA and BeppoSax are from Xia et al. (1998), 
Iwasawa (1999) and Risaliti et al. (2000), respectively. Notice that for
ASCA and BeppoSax observations only the total fluxes of Mrk~273 and Mrk~273x
are given as they do not resolve these two sources.}
\tablenotetext{b}{This flux is in the 0.5-2.0 keV band from Iwasawa (1999).}   
\enddata
\end{deluxetable}

\end{document}